\documentclass[twoside,12pt]{article}

\usepackage{indentfirst}
\usepackage{bm}
\usepackage{epsfig}
\usepackage[utf8]{inputenc}
\usepackage[colorlinks,allcolors=blue]{hyperref}
\usepackage[square,comma,sort&compress,numbers]{natbib}
\usepackage{CJK}
\usepackage{amsmath}

\begin{document}

\begin{CJK*}{UTF8}{}

\thispagestyle{empty} \vspace*{0.8cm}\hbox
to\textwidth{\vbox{\hfill\huge\sf Commun. Theor. Phys.\hfill}}
\par\noindent\rule[3mm]{\textwidth}{0.2pt}\hspace*{-\textwidth}\noindent
\rule[2.5mm]{\textwidth}{0.2pt}


\begin{center}
\LARGE\bf Effects of the tensor force on low-energy heavy-ion fusion reactions: A mini review$^{*}$
\end{center}

\footnotetext{\hspace*{-.45cm}\footnotesize $^\dag$Corresponding author, E-mail: luguo@ucas.ac.cn }

\begin{center}
\rm
Xiang-Xiang Sun \CJKfamily{gbsn}(孙向向)
and Lu Guo \CJKfamily{gbsn}(郭璐) $^\dagger$
\end{center}

\begin{center}
\begin{footnotesize} \sl
{School of Nuclear Science and Technology,
 University of Chinese Academy of Sciences,
 Beijing 100049, China}$^{\rm a)}$ \\
{CAS Key Laboratory of Theoretical Physics,
              Institute of Theoretical Physics,
              Chinese Academy of Sciences,
              Beijing 100190, China}$^{\rm b)}$ \\
\end{footnotesize}
\end{center}
\end{CJK*}

\begin{center}
\footnotesize (Received XXXX; revised manuscript received XXXX)

\end{center}

\vspace*{2mm}

\begin{center}
\begin{minipage}{15.5cm}
In recent several years, the tensor force, one of the most important components
of the nucleon-nucleon force, has been implemented in time-dependent density functional
theories and it has been found to influence many aspects of low-energy
heavy-ion reactions, such as dissipation dynamics, sub-barrier fusions,
low-lying vibration states of colliding partners.
Especially, the
effects of tensor force on fusion reactions have been
investigated from the internuclear potential to fusion cross sections systematically.
In this work we present a mini review on the recent progresses on this topic.
Considering the recent progress of low-energy reaction theories,
we will also mention
more possible effects of the tensor force on reaction dynamics.

\end{minipage}
\end{center}

\begin{center}
\begin{minipage}{15.5cm}
\begin{minipage}[t]{2.3cm}{\bf Keywords:}\end{minipage}
\begin{minipage}[t]{13.1cm}
Low-energy heavy-ion collision, time-dependent density functional theory,
the tensor force, internuclear potential, fusion cross sections
\end{minipage}\par\vglue8pt

\end{minipage}
\end{center}

\section{Introduction}
The tensor force is one of the important components of the nucleon-nucleon forces and is of great interests in nuclear physics. The most typical example to show the significance of the tensor force is that it provides a strong
central attraction in the isospin zero channel which is responsible for the deuteron binding.
With the development of radioactive-ion-beam facilities (RIBFs),
many exotic phenomena,
such as the existence of nuclear halos and skins \cite{Tanihata1985_PRL55-2676,Tanihata2013_PPNP68-215},
clustering effects \cite{Tohsaki2001_PRL87-192501},
shape coexistence \cite{Cwiok2005_Nature433-705},
the changes of shell closures \cite{Ozawa2000_PRL84-5493},
the pygmy resonances in electric dipole transitions \cite{Paar2007_RPP70-691}
have been observed in the region far away from the $\beta$-stability valley.
The studies of these exotic nuclear structures are at the forefront of nuclear research nowadays and many experimental and theoretical efforts have been made in recent years \cite{Bender2003_RMP75-121,Cwiok2005_Nature433-705,Meng2006_PPNP57-470,
Heyde2011_RMP83-1467,Meng2015_JPG42-093101,Niksic2011_PPNP66-519,Meng2016_RDFNS,
Zhou2016_PS91-063008,Zhou2017_PoS-INPC2016-373,Freer2018_RMP90-035004,
Otsuka2020_RMP92-015002}.
The tensor force plays an particularly important role in the shell evolution in exotic nuclei \cite{Otsuka2005_PRL95-232502}.
This has motivated many studies of the effect of the
tensor force on the shell structure of exotic nuclei,
spin-orbital splitting,
single-particle states, deformations by using nuclear density functional
theories or shell model.
Encouraging progresses have been made, see Refs. \cite{Sagawa2014_PPNP76-76,Otsuka2020_RMP92-015002} for recent reviews.
More interesting and complicated topics are
to explore the relation between nuclear structure
and reaction mechanism
\cite{Back2014_RMP86-317360,Jiang2021_EPJA57-23547},
especially via low-energy heavy-ion reactions of exotic nuclei,
such as weakly bound or halo nuclei
and nuclei with a large neutron excess,
and to understand the role of the tensor force
in these processes.

The microscopic description for the low-energy
heavy-ion collisions should originate from the underlying interaction, the interaction between the nucleons. 
The understanding of reaction dynamics is still one of the most challenging topics in nuclear physics nowadays.
The study of the heavy-ion reactions with
microscopic models and the comparison between results from model calculations and experimental data \cite{Negele1982_RMP54-913,Wen2013_PRL111-012501} are helpful to test some properties of the nuclear force and to reveal the influence of
nuclear structures on reactions.
At present, the microscopic description of reaction process can be achieved by using the time-dependent
density functional theories, in which
the adopted effective interactions between nucleons are constructed from basic symmetries of the nuclear force and the involved parameters are determined by fitting to
characteristic experimental data of finite nuclei and nuclear matter.
These effective
interactions include
non-relativistic and relativistic ones and
the time-dependent calculations based on both \cite{Ren2019_SciChinaPMA62-112062,Ren2020_PRC102-044603,
Ren2020_PLB801-135194,Umar2006_PRC73-054607,Maruhn2014_CPC185-2195} have been achieved.
It should be mentioned that in the time-dependent covariant density functional theory developed in recent years
\cite{Ren2019_SciChinaPMA62-112062,Ren2020_PRC102-044603,Ren2020_PLB801-135194,Ren2022_PRL128-172501},
the tensor force is not included.
Therefore in this work, we focus on the time-dependent calculations
by using the non-relativistic effective interactions, i.e., the Skyrme force,
to show the effect of tensor force on low-energy heavy-ion reactions.

In heavy-ion reactions, various couplings between
the relative motion and the excitation in the internal degrees of freedom of colliding systems should be considered in principle and thus make the description of the reaction mechanisms more complicated.
The internal excitations of reactants include collective ones and non-collective ones, such as low-lying vibrations
\cite{Morton1994_PRL72-4074,Stefanini1995_PRL74-864,Simenel2013_PRC88-064604},
nucleon(s) transfer
\cite{Washiyama2009_PRC80-031602R,Simenel2010_PRL105-192701,Sekizawa2017_PRC96-041601R},
rotations \cite{Leigh1995_PRC52-3151}, and
high-lying giant resonances \cite{Diaz-Torres2008_PRC78-064604}.
The coupling between the relation motion and internal excitation can be usually treated by using the coupled-channels approaches \cite{Hagino2022_PPNP-103951} and
it has been shown that
the reaction dynamics and, subsequently, the outcome of the
reaction can be affected by these couplings.

The effects of tensor force on these couplings in heavy-ion reaction dynamics are still open questions.
It has been mentioned that the low-energy heavy-ion reactions are affected by different couplings between relative motion and the excitation of collision partners.
Among these couplings, those from low-lying collective motion
and nucleons transfer, which strongly
influence near-barrier fusion,
are very sensitive to the underlying shell structure, which can be affected by the tensor force.
Another effect of the tensor force is that it modifies the dynamic dissipation in heavy-ion collisions \cite{Dai2014_SciChinaPMA57-1618}.
The latter has been the subject of theoretical studies at energies well above the Coulomb barrier~\cite{Dai2014_SciChinaPMA57-1618,
Stevenson2016_PRC93-054617,Shi2017_NPR34-41}.
In recent years, several heavy-ion fusion reactions between medium or heavy nuclei have been studied by the time-dependent density functional theory with considering the tensor force and
how the tensor force influences the fusion cross section has been investigated \cite{Stevenson2016_PRC93-054617,Guo2018_PLB782-401,Guo2018_PRC98-064607,
Godbey2019_PRC100-054612,Li2019_SCPMA62-122011,Sun2022_PRC105-034601}.
Therefore, in this work we focus on the recent progress on the study of fusion reactions
with considering the contributions from tensor force.

To study fusion reactions at both above and below-barrier energies,
the starting points of most theoretical approaches is
the nucleus-nucleus potential.
There are mainly two kinds of models to determine the nucleus-nucleus potential:
phenomenological potentials
\cite{Bass1974_NPA231-45,Randrup1978_PLB77-170,
Satchler1979_PR55-183,Adamian2004_PRC69-044601,
Swiatecki2005_PRC71-014602,Feng2006_NPA771-50,
Wang2012_PRC85-041601R,
Zhu2014_PRC89-024615,Zagrebaev2015_NPA944-257,
Bao2016_PRC93-044615,Wang2017_ADNDT114-281370}
and (semi)microscopic ones
\cite{Brueckner1968_PR173-944,
Diaz-Torres2001_NPA679-410,
Moeller2004_PRL92-072501,
Guo2004_NPA740-59,
Guo2005_PRC71-024315,
Umar2006_PRC74-021601R,
Wang2006_PRC74-044604,
Misicu2007_PRC75-034606,
Diaz-Torres2007_PLB652-255,
Washiyama2008_PRC78-024610,
Simenel2013_PRC88-024617,
Wen2013_PRL111-012501,
Simenel2017_PRC95-031601R}.
The phenomenological models have been widely
applied to study many aspects of reactions,
but its predictive power is limited due to several
adjustable parameters,
such as
the Bass model
\cite{Bass1974_NPA231-45},
the proximity potential
\cite{Seiwert1984_PRC29-477,Randrup1978_PLB77-170},
the double-folding potential
\cite{Satchler1979_PR55-183},
and driven potential from dinuclear system model
\cite{Adamian2004_PRC69-044601}.
The fusion process is particularly complex
and the cross section is affected by
many effects.
The predictions from a microscopic model
in the nucleonic degrees of freedom is more reliable and particularly a microscopic description can be connected with
the underlying nuclear shell structure and
dynamic effects of the reaction system relies on the adopted density functionals.
In such a way, it is also possible to check and analyze how the underlying effective interaction and its components, such as tensor force, affect the dynamic process and fusion cross sections.
After obtaining the internuclear potential based on microscopic effective interactions with or without tensor components, one can calculate the fusion cross sections by using the standard coupled-channels method \cite{Hagino1999_CPC123-143}.
In this way, the effect of tensor force on fusion reactions can be analyzed.
To this end, this review is organized as follows. We introduce the theoretical framework to determine the internuclear potential with the Skyrme effective interaction with or without tensor force
and fusion cross sections in Section \hyperref[Sec2]{2}. Section \hyperref[Sec3]{3} and \hyperref[Sec4]{4} present the influences of the tensor force on the internuclear potentials and fusion cross sections, respectively. The summary and perspective are shown in section \hyperref[Sec5]{5}.

\section{Theoretical framework\label{Sec2}}
In this section, we firstly show the tensor component in the Skyrme effective interactions. Then a brief introduction of the TDHF is given. Three approaches to obtain the internuclear potentials and the method
used to calculate the fusion cross section are also presented.

\subsection{Tensor force in Skyrme effective interaction}
Although the TDHF approach has been widely applied to low-energy heavy-ion reactions,
various assumptions and approximations that might
affect the TDHF results and lead to the incorrect reproduction of measurements have been employed in the past.
To remedy these problems considerable theoretical and computational efforts have been taken to improve numerical treatments
and density functionals.
An early discrepancy between TDHF predictions and measurements
\cite{Umar1986_PRL56-2793,Umar1989_PRC40-706}
had been solved by including the contributions
from spin-orbit interactions,
which is turned out to play an important role in reaction dynamics
\cite{Maruhn2006_PRC74-027601,Dai2014_PRC90-044609}.
In recent years with the developments of
high-performance computing equipment, TDHF calculations on a three-dimensional Cartesian grid without any symmetry restrictions have been achieved.
Additionally, the time-odd interactions, which have non-negligible contributions to heavy-ion collisions, have also been included
\cite{Umar2006_PRC73-054607}.
Recently the tensor force is also implemented
in the state-of-art TDHF calculations
and it can also affect the reaction dynamics
\cite{Stevenson2016_PRC93-054617,Guo2018_PLB782-401}.

In this paper, we focus on the effects of tensor force on heavy-ion reactions.
The tensor terms in the Skyrme effective interaction \cite{Skyrme1956_MP1-1043}
read
\begin{align}
\begin{split}
v_T&=\dfrac{t_\mathrm{e}}{2}\bigg\{\big[3({\sigma}_\mathrm{1}\cdot\bm{k}')({\sigma}_\mathrm{2}\cdot\bm{k}')
-({\sigma}_\mathrm{1}\cdot{\sigma}_\mathrm{2})\bm{k}'^{\mathrm{2}}\big]
\delta(\bm{r}_\mathrm{1}-\bm{r}_\mathrm{2})\\
&+\delta(\bm{r}_\mathrm{1}-\bm{r}_\mathrm{2})\big[3({\sigma}_\mathrm{1}\cdot\bm{k})({\sigma}_\mathrm{2}\cdot\bm{k})
-({\sigma}_\mathrm{1}\cdot{\sigma}_\mathrm{2})\bm{k}^\mathrm{2}\big]\bigg\}\\
&+t_\mathrm{o}\bigg\{3({\sigma}_\mathrm{1}\cdot\bm{k}')\delta(\bm{r}_\mathrm{1}-\bm{r}_\mathrm{2})({\sigma}_\mathrm{2}\cdot\bm{k})-({\sigma}_\mathrm{1}\cdot{\sigma}_\mathrm{2})\bm{k}'
\delta(\bm{r}_\mathrm{1}-\bm{r}_\mathrm{2})\bm{k}\bigg\},
\end{split}
\end{align}
where $t_\textrm{e}$ and $t_\textrm{o}$ are the strengths of triplet-even and
triplet-odd tensor interactions, respectively.

In TDHF theory, the energy of a nucleus is a functional of various densities and reads
\begin{equation}
E=\int \text{d}^3r  {\cal H}\left(\rho, \tau, {\bm{j}}, {\bm{s}}, {\bm{T}}, {\bm{F}}, J_{\mu\nu}, {\bm{r}}\right),
\label{Energy}
\end{equation}
where $\rho$, $\tau$, ${\bm{j}}$, ${\bm{s}}$, ${\bm{T}}$,  ${\bm{F}}$, and
$J$ are the number density, kinetic density, current density, spin density, spin-kinetic density,
the tensor-kinetic density, and spin-current pseudotensor density, respectively
\cite{Stevenson2016_PRC93-054617}.
Thus the full version of Skyrme energy density functional can be expressed as
\begin{align}
\label{EDFH}
\begin{split}
\mathcal{H}&=\mathcal{H}_0+\sum_{\rm{t=0,1}}\Big\{A_{\rm{t}}^{\rm{s}}\bm{s}_{\rm{t}}^2+
(A_{\rm{t}}^{\Delta{s}}+B_{\rm{t}}^{\Delta{s}})
\bm{s}_{\rm{t}}\cdot\Delta\bm{s}_{\rm{t}}+
B_{\rm{t}}^{\nabla s}(\nabla\cdot \bm{s}_{\rm{t}})^2 \\
&+B_{\rm{t}}^{F}\big(\bm{s}_{\rm{t}}\cdot
\bm{F}_{\rm{t}}-\frac{1}{2}\big(\sum_{\mu=x}^{z}J_{\rm{t}, \mu\mu}\big)^2-
\frac{1}{2}\sum_{\mu, \nu=x}^{z}J_{\rm{t}, \mu\nu}J_{\rm{t}, \nu \mu}\big)\\
&+(A_{\rm{t}}^{\rm{T}}+B_{\rm{t}}^{\rm{T}})\big(\bm{s}_{\rm{t}}\cdot\bm{T}_{\rm{t}}-
\sum_{\mu,\nu=x}^{z}J_{\rm{t},\mu\nu}J_{\rm{t},\mu\nu}\big)\Big\},
\end{split}
\end{align}
where $\mathcal{H}_0$ is the simplified functional used in the TDHF code
\texttt{Sky3D} \cite{Maruhn2014_CPC185-2195} and most TDHF calculations.
Those terms with the coupling constants $A$ come from the central
and spin-orbit interactions and those with $B$ from the tensor force.
The details of $A$ and $B$
can be found in Refs.~\cite{Lesinski2007_PRC76-014312,Davesne2009_PRC80-024314}.
It should be mentioned that up to now all the time-even and time-odd terms in Eq.~(\ref{EDFH}) have been included in the static Hartree-Fock (HF), time-dependent Hartree-Fock (TDHF),
and density-constrained (DC) TDHF calculations.

Two ways have been applied to determine the parameterization of the tensor components in Skyrme density functionals.
The first one is to include the tensor force perturbatively to the existing density functionals,
for instance,
the effective interaction SLy5 \cite{Chabanat1998_NPA635-231} plus tensor force, labeled as SLy5t~\cite{Colo2007_PLB646-227}.
Therefore, by making the comparison between calculations with SLy5 and those with SLy5t,
one can know the changes caused by the tensor force itself.
By readjusting the full set of Skyrme parameters self-consistently, the strength of the tensor force can also be determined.
This strategy has been used in Ref.~\cite{Lesinski2007_PRC76-014312} and led to
dozens of parameter sets of tensor interactions,
denoted as T$IJ$.
Due to its fitting strategy, the contributions from the tensor force and the rearrangement of all other terms are physically entangled.

\subsection{TDHF approach}

The action corresponding to a given Hamiltonian can be constructed as
\begin{equation}
S=\int_{t_1}^{t_2} \text{d}t \langle \Phi(\bm{r}, t)|H-i\hbar \partial_t|\Phi(\bm{r}, t)\rangle ,
\end{equation}
where $\Phi(\bm{r}, t)$ is the time-dependent wave function for the many-body system
with $N$ nucleons.
Under the mean-field approximation,
the many-body wave function $\Phi(\bm{r}, t)$ is the single time-dependent Slater determinant constructed by the single particle wave functions $\phi_{\rm{\lambda}}(\bm{r}, t)$ and reads
\begin{equation}
\Phi(\bm{r}, t)=\frac{1}{\sqrt{N!}}\textrm{det}\{\phi_{\rm{\lambda}}(\bm{r}, t)\}.
\end{equation}
With the variation principle, i.e.,
taking the variation of the action with respect to the single-particle states,
the equations of motion of the $N$ nucleons are
\begin{equation}
i\hbar \frac{\partial}{\partial_t}\phi_{\rm{\lambda}}(\bm{r}, t)=h\phi_{\rm{\lambda}}(\bm{r}, t), \quad \lambda = 1, \cdots, N,
\end{equation}
with the single-particle Hamiltonian $h$.
These nonlinear TDHF equations have been solved accurately on three-dimensional
coordinate space without any symmetry restrictions \cite{Umar2006_PRC73-054607,Maruhn2014_CPC185-2195}.
The TDHF approach can provide a microscopic description of nuclear dynamics, as seen in recent applications to fusion reactions
\cite{Simenel2004_PRL93-102701,Umar2009_PRC80-041601R,
Oberacker2010_PRC82-034603,Guo2012_EWC38-09003,
Keser2012_PRC85-044606,Umar2012_PRC85-055801,
Simenel2013_PRC88-024617,Umar2014_PRC89-034611,
Washiyama2015_PRC91-064607,Tohyama2016_PRC93-034607,
Godbey2017_PRC95-011601R,Simenel2017_PRC95-031601R,Sun2022_PRC105-054610},
quasifission process
\cite{Golabek2009_PRL103-042701,Oberacker2014_PRC90-054605,
Umar2015_PRC92-024621,Umar2016_PRC94-024605,
Yu2017_SCPMA60-0920116},
transfer reactions
\cite{Washiyama2009_PRC80-031602R,Simenel2010_PRL105-192701,
Simenel2011_PRL106-112502,Scamps2013_PRC87-014605,
Sekizawa2013_PRC88-014614,Wang2016_PLB760-236,
Sekizawa2016_PRC93-054616,Sekizawa2017_PRC96-041601R},
fission
\cite{Simenel2014_PRC89-031601R,Scamps2015_PRC92-011602R,
Goddard2015_PRC92-054610,Goddard2016_PRC93-014620,
Bulgac2016_PRL116-122504,Tanimura2017_PRL118-152501,
Bulgac2022_PRL128-022501,Bulgac2019_PRC100-034615,
Bulgac2021_PRL126-142502},
and deep inelastic collisions
\cite{Maruhn2006_PRC74-027601,Guo2007_PRC76-014601,
Guo2008_PRC77-041301R,Iwata2011_PRC84-014616,
Dai2014_PRC90-044609,Dai2014_SciChinaPMA57-1618,
Stevenson2016_PRC93-054617,Guo2017_EWC163-00021,
Shi2017_NPR34-41,Umar2017_PRC96-024625}.
More applications of the TDHF can be found in recent reviews
\cite{Simenel2012_EPJA48-152,
Nakatsukasa2016_RMP88-045004,
Simenel2018_PPNP103-19,
Stevenson2019_PPNP104-142164,
Sekizawa2019_FP7-20}.

\subsection{Microscopic internuclear potentials and fusion cross sections}
The TDHF simulation, based on the mean-field approximation,
provides the most probable trajectory
of the collision system and the quantum fluctuation is not
included.
The quantum tunneling of the many-body wave function
cannot be treated with the TDHF approach.
As a consequence,
the TDHF approach cannot be
directly applied to study sub-barrier
fusion reactions.
It should be noted that an imaginary-time mean-field method might be
the answer of this problem \cite{McGlynn2020_PRC102-064614}.
The fusion cross section can be estimated by the quantum sharp-cutoff formula \cite{Bonche1978_PRC17-1700,Umar2006_PRC73-054607,Guo2018_PRC98-064609,Li2019_SCPMA62-122011} based on a mass of TDHF simulations with different incident parameters or angular momenta,
but which may underestimate
the cross sections especially for sub-barrier collisions.
Currently,
all approaches to study sub-barrier fusions assume that there is an internuclear potential
which depends on the internuclear distance
and the fusion reaction is usually treated as
a quantum tunneling through this potential
in the center-of-mass frame.
The internuclear potential can also be calculated microscopically
with the TDHF approach by applying
frozen Hartree-Fock (FHF)~\cite{Guo2012_EWC38-09003,Bourgin2016_PRC93-034604},
density-constrained FHF (DCFHF)~\cite{Simenel2017_PRC95-031601R,Umar2021_PRC104-034619},
DC-TDHF~\cite{Umar2006_PRC74-021601R}, or
dissipative-dynamics TDHF~\cite{Washiyama2008_PRC78-024610} approach.
The obtained potential can be used to calculate
penetration probabilities with the incoming wave boundary
condition method
\cite{Hagino1999_CPC123-143}.
In the section, we will introduce the FHF, DCFHF, and DC-TDHF method
and how to calculate the cross section with those internuclear potentials.

Based on the TDHF dynamic evolution of the collision system, the
internuclear potential can be extracted by using the density constraint technique.
As a consequence, the obtained potential contains all dynamic effects,
such as neck formation, dynamic deformation effects, and particle transfer.
In this approach, at certain moment during the TDHF evolution,
the instantaneous TDHF density is used to perform a static HF energy minimization
\begin{equation}
\delta \langle \Psi_{\rm DC}|H-\int \text{d}^3r \lambda(\textbf{r})\rho(\textbf{r})|\Psi_{\rm DC}\rangle=0,
\end{equation}
by constraining the proton and neutron densities to be the same as the instantaneous TDHF densities.
Since the total density are constrained to be unchanged,
all mass moments are simultaneously constrained.
$\lambda$ is the Lagrange parameter at each point of space.
$|\Psi_{\rm DC}\rangle$ is the many-body wave function under the density constraint.
The energy corresponding to the $|\Psi_{\rm DC}\rangle$, i.e., the density-constrained energy, reads
\begin{equation}
E_{\rm {DC}}(\bm R)=\langle \Psi_{\rm DC}|H|\Psi_{\rm DC}\rangle.
\end{equation}
This energy still includes the binding energies of two colliding nuclei,
which should be substracted.
Then the internuclear potential is given by
\begin{equation}
V_\mathrm{DC-TDHF}(\bm R)=E_{\rm {DC}}(\bm R)-E_{\rm {P}}-E_{\rm {T}},
\label{VB}
\end{equation}
where $E_{\rm {P}}$ and $E_{\rm {T}}$ are the binding energies of
the projectile ($\mathrm{P}$) and target ($\mathrm{T}$), respectively.
It should be noted that the density-constraint procedure does not influence
the TDHF evolution and not contain any free parameters or normalization.
In this approach, all the single-particle levels are allowed to reorganize during
minimizing the total energy, thus the Pauli exclusion principle is included dynamically.

Comparing with the internuclear potentials from DC-TDHF,
those from the DCFHF method does not include any dynamic factors
and the contribution from Pauli exclusion principle is still included.
The Pauli exclusion principle is included by allowing the single particle states
to reorganize to attain minimum energy 
in the static HF calculations with the density constraint and to
be properly antisymmetrized, as the many-body state is a
Slater determinant of all the occupied single-particle wave
functions.
The HF calculations are preformed with constraining the
the total proton $p$ and neutron $n$ densities to be
the same as those at the ground state,
\begin{equation}
	\delta \left \langle
	H-\int \text{d}^3r \sum_{q=p,n} \lambda_q(\textbf{r})
	\left[\rho_{\rm{P},q} (\bm r) +
	\rho_{\rm{T},q} (\bm r- \bm R)
	\right]\right\rangle=0,
\end{equation}
where $\rho_{\rm{P}}$ and $\rho_{\rm{T}}$ are the densities of
the projectile and target in their ground states. This variation procedure results in a unique
Slater determinant $\Phi({\bm R})$. Similar to the case of DC-TDHF,
the internuclear potential from DCFHF is given by
\begin{equation}
	V_\mathrm{DCFHF}(\bm R)=\langle \Phi({\bm R})|H|\Phi({\bm R})\rangle -E_{\rm {P}}-E_{\rm {T}}.
	\label{VB1}
\end{equation}

From above, the DC-TDHF has been introduced to compute the nucleus-nucleus potential in a dynamical microscopic way.
All of the dynamical effects included in TDHF
and Pauli exclusion principle are then directly incorporated.
For the one from DCFHF, the dynamic effects are not included while
the Pauli exclusion principle is still kept.
There is also a potential
which includes neither dynamical effects nor the contributions of Pauli exclusion principle.
This nucleus-nucleus potential is defined as the potential between the nuclei in their ground states.
This is achieved with the frozen Hartree-Fock (FHF) technique \cite{Simenel2013_PRC88-064604},
assuming that the densities of the target and
projectile are unchanged and equal to their ground state densities.
The potential can then be expressed as
\begin{align}
\label{eq:FD}
V_\mathrm{FHF}(\bm R)=E[\rho_\mathrm{P}+\rho_\mathrm{T}](\bm R)-E[\rho_\mathrm{P}]-E[\rho_\mathrm{T}].
\end{align}
When the internuclear distance is large, i.e., the overlap between the densities of the projectile and target is small, the Pauli principle
almost has no influence on the internuclear potential.
However, when two nuclei are close to each other and the density overlaps are large, the Pauli principle is expected to play an important role.
Therefore the FHF approximation can not properly
describe the inner part of the potential
\cite{Simenel2017_PRC95-031601R}.

After obtaining internuclear potentials, the fusion cross section can be calculated.
It has been mentioned that including the couplings between the collective excitation of the target and
projectile plays a significant role to describe fusion excitation function.
As mentioned above, the internuclear potentials from FHF and DCFHF do not include any dynamic information and can be directly applied to coupled-channels approach \cite{Hagino1999_CPC123-143} to calculate the fusion cross sections \cite{Simenel2017_PRC95-031601R}.
It should be noted that the combination of FHF method with the coupled-channels calculation can provides a reasonable description of fusions at near-barrier regions \cite{Simenel2013_PRC88-064604}, but cannot be used to study fusion reactions well below the barrier or in systems with large $Z_\mathrm{P}Z_\mathrm{T}$ \cite{Dasso2003_PRC68-054604,Umar2021_PRC104-034619} because the Pauli repulsion is not included.
When calculating the fusion cross sections by using the potentials from DCFHF together with the coupled-channels method, it has been shown that calculated results are more consistent with the measurements than those from FHF potential \cite{Simenel2017_PRC95-031601R}, especially in sub-barrier region.

The DC-TDHF approach provides a microscopic way to obtain internuclear potentials, which already contains all the dynamic factors and is connected with the coordinate-dependent mass $M(R)$.
Therefore, the fusion cross sections can be directly calculated by solving the
one dimension
Schr\"odinger equation with the potential extracted from DC-TDHF.
The procedure to obtain transmission probabilities $T_{l}(E_{\mathrm{c.m.}})$ (and thus cross sections) from an arbitrary one-dimensional potential $V(R)$ can be calculated by solving the Schr\"odinger equation
\begin{equation}
\left[\frac{-\hbar^2}{2{M(R)}}\frac{\text{d}^2}{\text{d}R^2}+\frac{l(l+1)\hbar^2}{2{M(R)}R^2} + V(R) - E\right]\psi=0,
\label{eq:se}
\end{equation}
where $l$ is the angular momentum of each partial wave.
Generally, the incoming wave boundary conditions (IWBC) method is used to calculate $T_{l}(E_{\mathrm{c.m.}})$ with the assumption that fusion occurs once the minimum of $V(R)$ is reached.
After obtaining $T_{l}(E_{\mathrm{c.m.}})$, fusion cross sections
are given by
\begin{equation}
\sigma_f(E_{\mathrm{c.m.}})=\frac{\pi}{k_0^2}\sum_{l=0}^{\infty}(2l+1)T_l(E_{\mathrm{c.m.}}).
\label{Eq:sigma}
\end{equation}

Since DC-TDHF potentials are the results of the TDHF evolution, the coordinate-dependent mass $M(R)$ can
be calculated with the energy conversation condition \cite{Umar2009_PRC80-041601R,Umar2009_EPJA39-243247}
\begin{equation}
	M(R)=\frac{2[E_\mathrm{c.m.}-V(R)]}{\dot{R}^2}.
\end{equation}
This coordinate-dependent mass mainly influences the inner part of the potential, leading to a broader
barrier width thus further suppressing the fusion cross sections at the sub-barrier region.
The potentials from DC-TDHF are dependent on the incident energy
$E_\mathrm{c.m.}$ and the energy-dependence behavior is also affected by
the coordinate-dependent mass \cite{Umar2014_PRC89-034611}.
Instead of solving the Schr\"odinger equation using the coordinate-dependent mass $M(R)$,
one can also calculated the fusion cross sections by using
a transformed potential \cite{Umar2009_EPJA39-243247,Goeke1983_AP150-504} and the scale transformation reads
\begin{equation}
d\bar{R}=\left(\frac{{M(R)}}{\mu}\right)^{\frac{1}{2}}dR.
\end{equation}
After this transformation the coordinate-dependence of $M(R)$ is replaced by the reduced mass $\mu$ in Eq.~(\ref{eq:se}) and the Schr\"odinger equation is solved by using the modified Numerov method
with the transformed potential. The details have been introduced in the coupled-channels code \texttt{CCFULL}~\cite{Hagino1999_CPC123-143}.
With the internuclear potentials from DC-TDHF approach,
the fusion cross sections at below and above-barrier energies of many systems are studies and good agreements between calculations and experimental data are achieved
\cite{Umar2006_PRC74-021601R,Umar2006_PRC74-024606,
Umar2008_PRC77-064605,Umar2009_EPJA39-243247,
Umar2009_PRC80-041601R,Umar2010_PRL104-212503,
Oberacker2010_PRC82-034603,Keser2012_PRC85-044606,
Umar2012_PRC85-055801,Simenel2013_PRC88-024617,
Oberacker2013_PRC87-034611,Umar2014_PRC89-034611,
Umar2015_NPA944-238,Umar2016_PRC94-024605,
Guo2018_PRC98-064607,Guo2018_PLB782-401,
Godbey2019_PRC100-054612,Godbey2019_PRC100-024619,
Sun2022_PRC105-034601}.

\section{Effects of tensor force on internuclear potentials}
\label{Sec3}
\begin{figure}
\begin{center}
\includegraphics[width=0.45\textwidth]{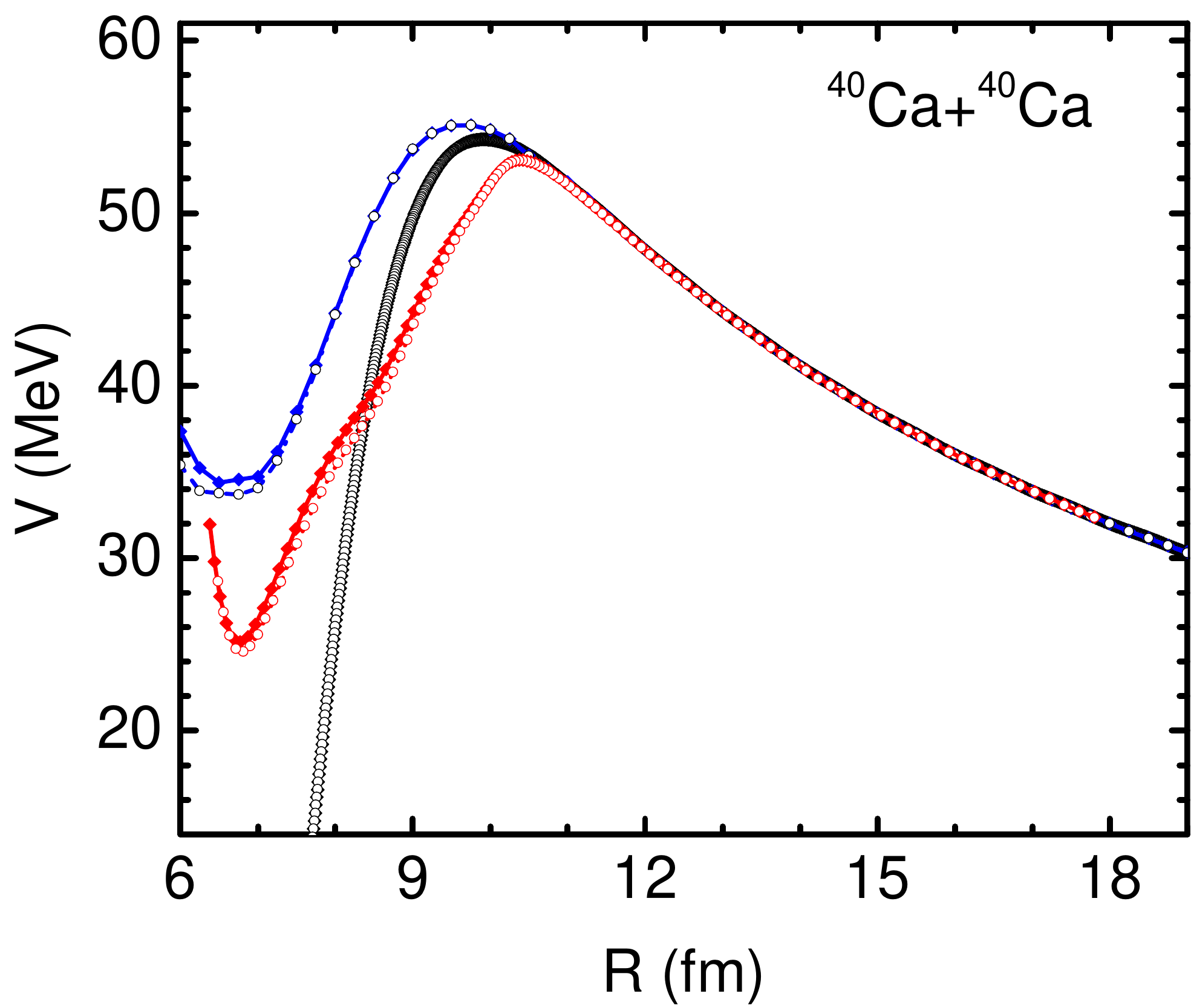}
\includegraphics[width=0.45\textwidth]{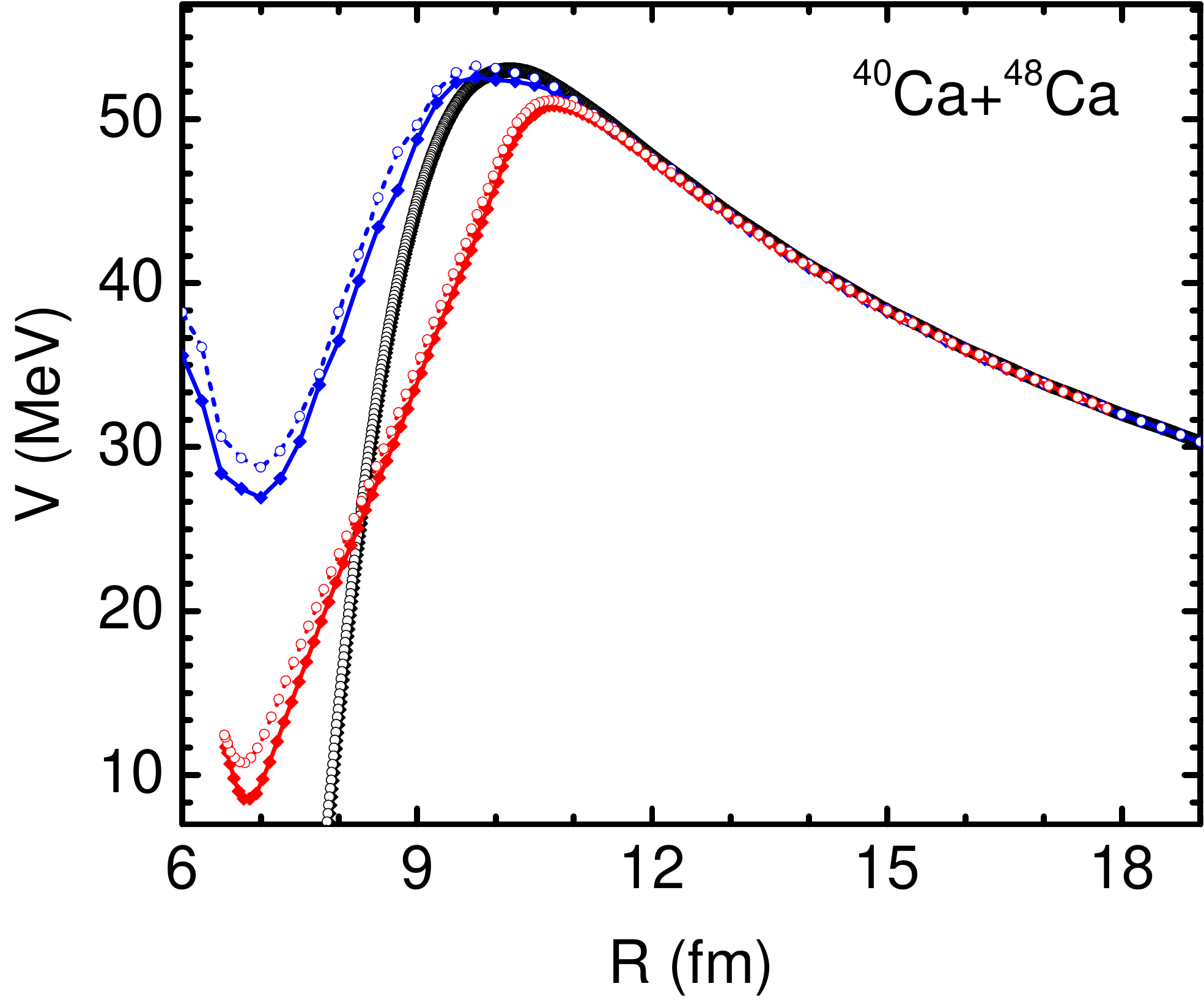}
\includegraphics[width=0.45\textwidth]{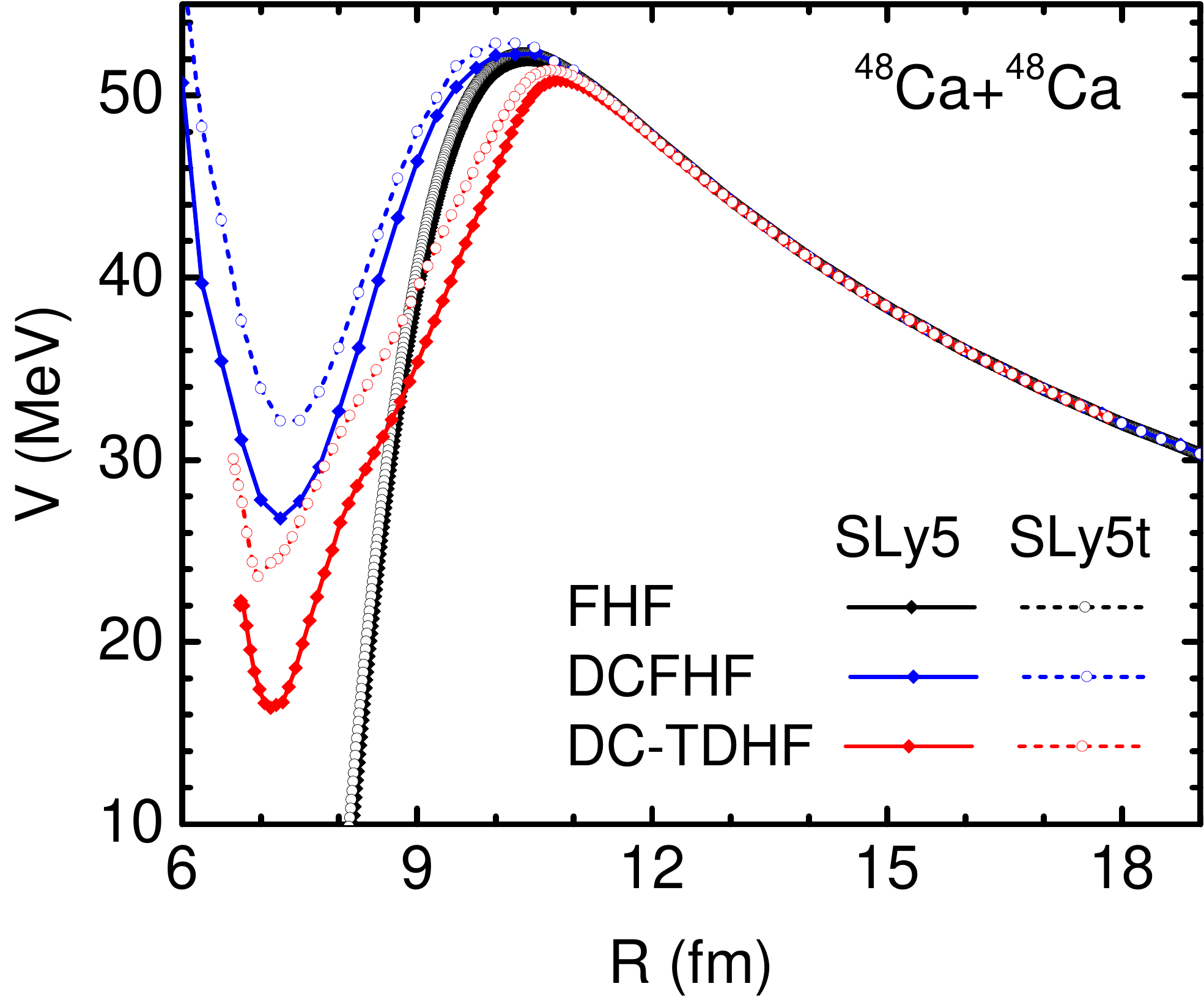}
\end{center}
\caption{
\label{fig:CaCa}
Internuclear potentials for $^{40}\mathrm{Ca}+{}^{40}\mathrm{Ca}$,
$^{40}\mathrm{Ca}+{}^{48}\mathrm{Ca}$, and
$^{48}\mathrm{Ca}+{}^{48}\mathrm{Ca}$
by using FHF, DCFHF, and DC-TDHF methods with the effective interactions SLy5 and SLy5t.
The results from FHF and DC-TDHF calculations are taken from Ref. \cite{Guo2018_PRC98-064607}. }
\end{figure}

The internuclear potentials from FHF, DCFHF, and DC-TDHF methods only rely on the adopted effective interactions and there is no additionally readjusted parameters.
The influence of tensor force can be examined by making a comparison between the calculated potentials with the tensor force and those without the tensor force.
In Fig. \ref{fig:CaCa}, we show the internuclear potentials between  $^{40}\mathrm{Ca}+{}^{40}\mathrm{Ca}$,
$^{40}\mathrm{Ca}+{}^{48}\mathrm{Ca}$, and
$^{48}\mathrm{Ca}+{}^{48}\mathrm{Ca}$
by using FHF, DCFHF, and DC-TDHF methods with the effective interactions SLy5 and SLy5t.
Before discussing the effects of the tensor force, we make a comparison among these three types of
potential.
It is found that for a reaction system with given effective interaction, the internuclear potentials from DC-FHF have the highest capture barriers while barriers corresponding to the DC-TDHF method are lowest.
Due to the absence of the Pauli repulsion, the internuclear potentials from the FHF method
do not have the strong repulsion core.
The inclusion of the Pauli repulsion increases the
height of Coulomb barrier, leads to the appearance
of the potential pocket in the inner part, and changes the shape of the barrier. When including dynamic effects, i.e., for the potentials from DC-TDHF calculations, both the minimum the potential pocket and the capture barriers become lower because of the inclusion of dynamic factors.

In Ref. \cite{Guo2018_PRC98-064607}, different combinations of projectile and target with proton and neutron numbers being the magic numbers 8, 20, 28, 50, and 82 have been chosen to investigate how the tensor force affects the nucleus-nucleus potential from FHF and DC-TDHF calculations.
Among these magic numbers, the spin-saturated shells are 8 and 20.
It is found that for light systems or reactants with spin-saturated shells the
tensor force slightly affects the barrier height and inner part of the barrier by a fraction of an MeV and for medium and heavy spin-unsaturated reactions the effect is much more obvious, with changes from a fraction of an
MeV to almost 2 MeV.
Figure~\ref{fig:CaCa} shows the comparison of the internuclear potentials obtained from the three methods mentioned-above with density functionals
SLy5 and SLy5t.  It is found that for $^{40}\mathrm{Ca}+{}^{40}\mathrm{Ca}$, the tensor almost has no influence on internuclear potential because the shell closure at $N(Z)=20$ is spin-saturated. For $^{40}\mathrm{Ca}+{}^{48}\mathrm{Ca}$ and $^{48}\mathrm{Ca}+{}^{48}\mathrm{Ca}$, it is found that the tensor force increases the capture barriers and minimum of potential pockets.
Additionally, for the internuclear potentials for $^{40,48}$Ca+$^{78}$Ni with SIII(T) \cite{Beiner1975_NPA238-29,Brink2018_PRC97-064304} and SLy5(t) shown in Ref. \cite{Sun2022_PRC105-034601}, it is found that the tensor can influence the shape of barrier.

\begin{figure}
\begin{center}
\includegraphics[width=0.95\textwidth]{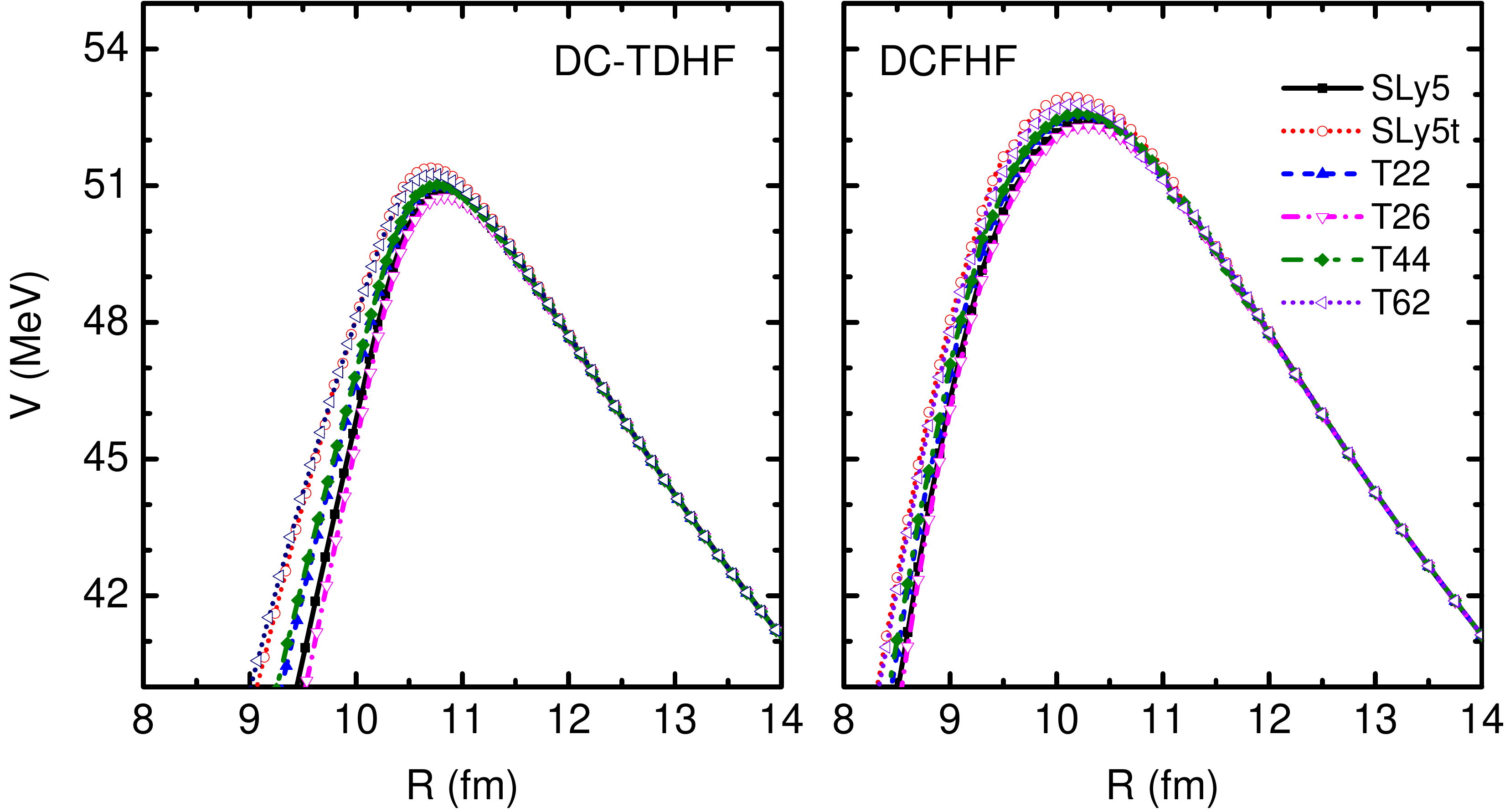}
\end{center}
\caption{
\label{fig:Ca48_TIJ}
Internuclear potentials for $^{48}\mathrm{Ca}+{}^{48}\mathrm{Ca}$
by using DC-TDHF (left panel) and DCFHF (right panel) with effective interactions SLy5, SLy5t, T22, T24, T44, and T62. }
\end{figure}

\begin{figure}
\begin{center}
\includegraphics[width=0.5\textwidth]{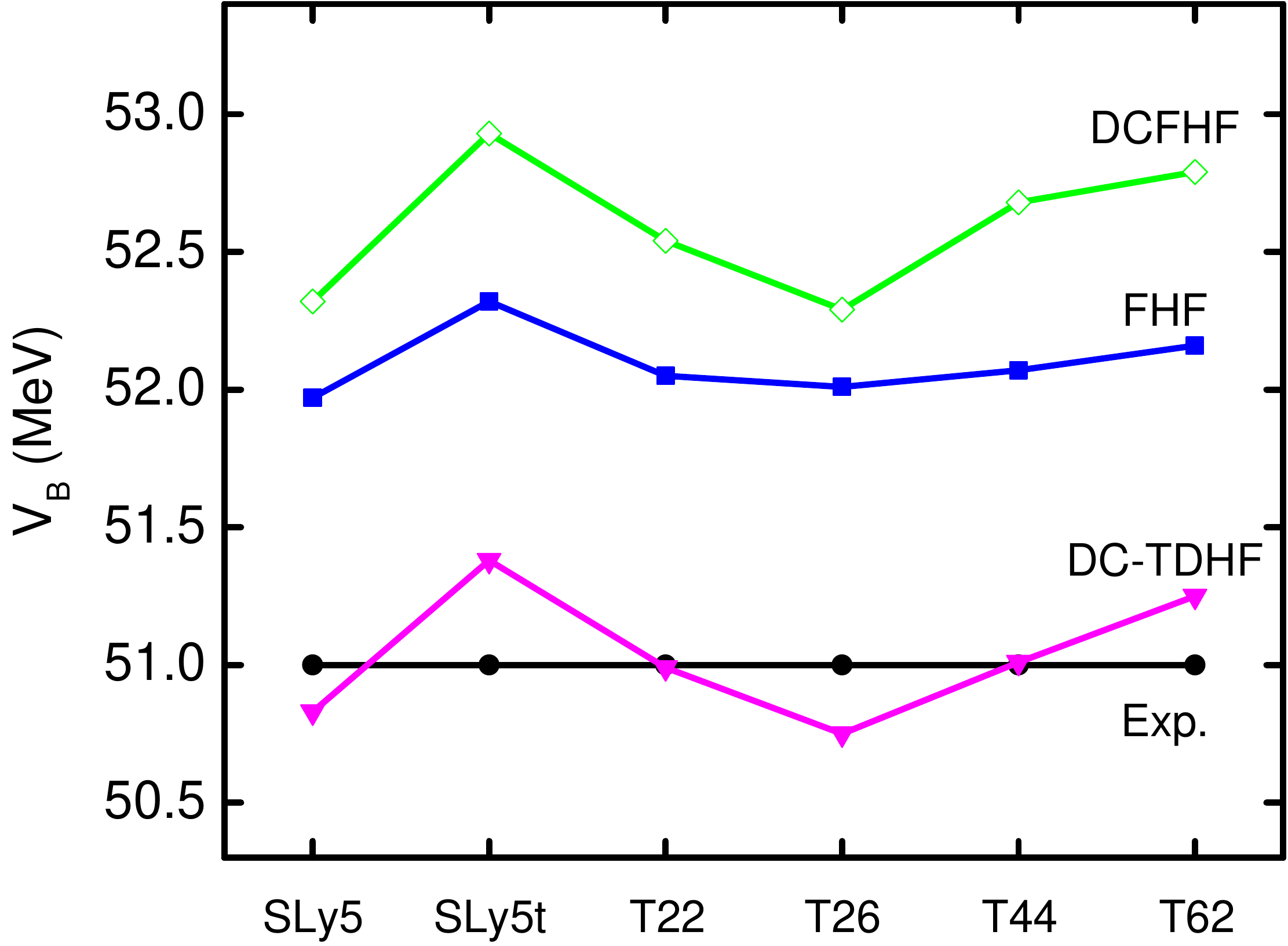}
\end{center}
\caption{
\label{fig:VB_Ca48}
The heights of barriers for $^{48}\mathrm{Ca}+{}^{48}\mathrm{Ca}$ calculated by using FHF, DCFHF, and DC-TDHF with six Skyrme
forces as well as the experimental value taken from Ref. \cite{Stefanini2009_PLB679-95}.
The capture barriers for FHF and DC-TDHF calculations are extracted from the
results given in Ref. \cite{Guo2018_PRC98-064607}.  }
\end{figure}

As mentioned before, the tensor components in SLy5t force are added perturbatively.
Therefore, it is necessary to make a
comparison among the results of various forces, for which the coupling constants of tensor force
are obtained by readjusting the full set of Skyrme parameters self-consistently.
Taking the system $^{48}\mathrm{Ca}+\mathrm{^{48}Ca}$ as an example,
we show the nucleus-nucleus potential with the six forces SLy5, SLy5t, T22, T26, T44, and T62 in Fig.~\ref{fig:Ca48_TIJ} by using the DCFHF and DC-TDHF methods.
The height of barriers of these potentials
are shown in Fig. \ref{fig:VB_Ca48} and
the experimental capture barrier taken from Ref \cite{Stefanini2009_PLB679-95} is also given.
Generally speaking,
the barriers of potentials from DCFHF are higher and wider than those
from DC-TDHF method.
From Fig. \ref{fig:VB_Ca48}, it is clear that the barriers from DC-TDHF calculations are well consistent with the experiment and those of FHF and DCFHF are higher than the experiments.
This is due to the potentials from DC-TDHF contain dynamic effects, such as
dynamic deformations and neutron transfers.
It has been shown \cite{Guo2018_PRC98-064607} that
different parameterizations of tensor force
have different influences on the low-lying
excitations and neutron (proton) transfer
such that the effects on the barrier heights
differ by effective interaction.
In Ref. \cite{Guo2018_PRC98-064607},
the comparison among the internuclear potentials from DC-TDHF and related
discussions have been made.
The comparison between T22 and T44 indicates that the isoscalar channel has
negligible effect in this reaction.
By comparing the results with T$26$, T$44$, and T$62$,
one can see that the potential increases as the isovector tensor
coupling decreases.
The potentials with T$62$ (T$26$) have similar potentials as SLy5t (SLy5),
even though tensor coupling constants of them are quite different,
because the rearrangement of the mean-field for T$62$ (T$26$)
might cancel part of the tensor force in SLy5t (SLy5).

\section{Effects of tensor force on fusion cross sections}
\label{Sec4}
In Section \hyperref[Sec3]{3}, we have shown that the tensor
force not only influence the barrier height but also affect the shape
of barrier. Therefore, the tensor force should, in principle, affect
the fusion cross sections, particularly for sub-barrier energies.
To calculate the fusion cross section, as for the potentials from FHF
and DCFHF, one should fit them to the Woods-Saxon type potential and then uses
the \texttt{CCFULL} \cite{Hagino1999_CPC123-143} to calculate the fusion cross sections
with considering the coupling from low-lying excited states
\cite{Simenel2017_PRC95-031601R}.
As for the potential from DC-TDHF calculation, one can directly calculate
the penetration probabilities under the transformed potentials
such that the fusion cross sections can be obtained [cf. Eq.~(\ref{Eq:sigma})].

\begin{figure}
\begin{center}
\includegraphics[width=0.6\textwidth]{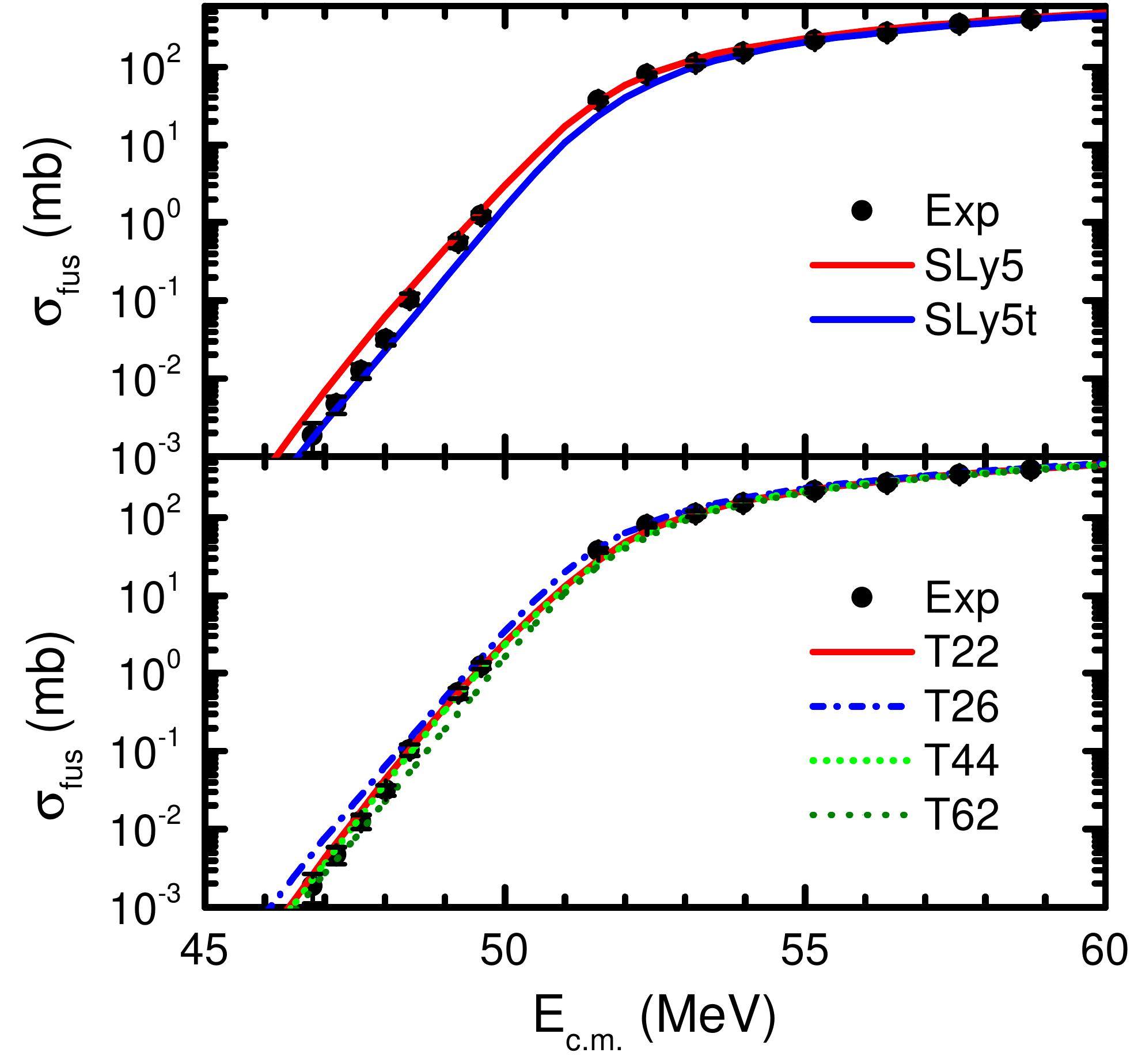}
\end{center}
\caption{
\label{fig:Ca48_sigma}
Fusion cross sections obtained by using the potentials from DCFHF calculations and experimental data taken from Ref. \cite{Stefanini2009_PLB679-95} for $^{48}\mathrm{Ca}+{}^{48}\mathrm{Ca}$.
The top panel shows the results calculated with the
effective interactions SLy5 and SLy5t. Those for effective interactions T22, T24, T44, and T62 are
shown in the bottom panel. }
\end{figure}

The tensor force almost has no influence on the internuclear potentials
of the system in which both the projectile
and target have spin-saturated neutron and proton shells, e.g.,
$^{40}\mathrm{Ca}+{}^{40}\mathrm{Ca}$.
In Ref. \cite{Godbey2019_PRC100-054612},
the effect of tensor on
the fusion cross sections obtained by using the potential from
DC-TDHF have been discussed and it is shown that for those systems
with spin-saturated shells,
the tensor force has no influence on fusion cross sections.
For the systems with spin-unsaturated proton (neutron) shells, the tensor force
has non-negligible effects, such as $^{40,48}\mathrm{Ca}+{}^{48}\mathrm{Ca}$
\cite{Godbey2019_PRC100-054612} and $^{40,48}\mathrm{Ca}+{}^{78}\mathrm{Ni}$
\cite{Sun2022_PRC105-034601}.
Especially for the sub-barrier region, the inclusion of the tensor force improves the
description of fusion cross sections for $^{48}\mathrm{Ca}+{}^{48}\mathrm{Ca}$
collision.
In Ref. \cite{Godbey2019_PRC100-054612}, a systematic comparison between the fusion cross sections
with considering the tensor force and those without the tensor force have been made.
Up to now, the effects of tensor force on fusion cross sections
by using the potential from DCFHF have not been clarified.
Therefore, we use the internuclear potentials from DCFHF to investigate this problem
by taking $^{48}\mathrm{Ca}+{}^{48}\mathrm{Ca}$
as an example.

The calculated capture cross sections with the potentials from DC-TDHF
has been shown in Ref. \cite{Godbey2019_PRC100-054612} and the results reproduce the data well.
In Fig.~\ref{fig:Ca48_sigma}, we show the fusion cross sections of $^{48}\mathrm{Ca}+{}^{48}\mathrm{Ca}$
calculated by using the code \texttt{CCFULL} with the
internuclear potentials from the DCFHF approach
and considering the coupling from low-lying vibration states $2^+_1$ and $3^-_1$ of $^{48}$Ca.
It is found that the calculations reproduce the
recent measurements \cite{Stefanini2009_PLB679-95} well.
The comparison between the results from SLy5 and SLy5t indicates that by including the tensor force perturbatively, the tensor force suppresses the fusion cross sections in the below-barrier region.
By comparing the results from DC-TDHF \cite{Godbey2019_PRC100-054612} and DCFHF,
one can find that the suppression of the cross section at sub-barrier region in DC-TDHF calculations
is more obvious than that of DCFHF. This is due to the heights of barriers corresponding to DCFHF method
are higher those of DC-TDHF.
After considering the tensor force, the description of the sub-barrier fusion cross section
is slightly improved.
For the results from $TIJ$ forces differ with each other,
suggesting the competition between isoscalar and isovector channels
plays a role in reaction dynamics.
For $TIJ$ forces, the conclusions obtained from DC-TDHF calculations also hold in DCFHF calculations.
It should be noted that here we directly use the experimental information of $2^+_1$ and $3^-_1$
given in Ref. \cite{Stefanini2009_PLB679-95} and it has been
shown in Ref. \cite{Guo2018_PLB782-401} the tensor force also influences the low-lying vibration properties of
exotic nuclei such that the influence of tensor force might be more complex.
A self-consistent and systematic studies on the influence of tensor force on low-lying spectra,
internuclear potentials, and fusion cross section are necessary.

\section{Summary and perspective}
\label{Sec5}

The effects of the tensor force on nuclear structures and nuclear reactions
are of great interests nowadays.
Accurate description of the fusion cross sections especially at the sub-barrier
energies for reactions involving exotic nuclei are very important topics in
nuclear reactions.
In this contribution, we briefly review recent progresses on the effects of
tensor force on low-energy heavy-ion fusion reactions.
We calculate the internuclear potentials for $^{40,48}$Ca+$^{40,48}$Ca
by using FHF, DCFHF, DC-TDHF methods and make a comparison among these
different internuclear potentials.
It is found that for spin-unsaturated systems, the tensor force has
considerable modifications on the inner part of internuclear potential
such that it affects the fusion cross sections, especially for sub-barrier collisions.

Although the effects of tensor force on fusion cross sections have been revealed,
there are still many open problems on the influence of tensor force on heavy-ion reaction
to be explored further.
Here we discuss part of them as the perspective of this work.
\begin{itemize}
\item In nuclear structure studies, it has been shown that the tensor force has strong influence
on the shell evolutions of exotic nuclei.
The shell structures are also very important for the studies on
nuclear reactions, especially on fission and quasifission processes.
Therefore it is particularly interesting and meaningful to investigate
how the tensor affect the (quasi) fission processes, the corresponding
fragments distributions, and the role of shell structure.
Such kinds of studies are under progress in our group.
  \item Up to now, the influences of tensor force on fusion reaction are
only studied in those systems with spin-(un)saturated shells
due to the absence of pairing effects.
Recently, fusion reactions with exotic nuclei is of great interests both
experimentally and theoretically.
It is necessary and meaningful to extend FHF, DCFHF, and DC-TDHF to study reactions with neutron(proton)-rich nuclei
with considering pairing correlations properly,
aiming at not only the precise description of fusion reaction with exotic nuclei,
but also a microscopic understanding of their reaction mechanism.
\item Recently, there are many experimental and theoretical
works concerning on the hindrance of the fusion cross sections
at the extreme sub-barrier energies.
One of the solution proposed to explain this phenomenon is
called the sudden approximation,
in which the hindrance of the fusion cross sections
is connected to the repulsive core of the internuclear potential.
We have shown that the tensor force have a strong influence on
the potential pocket.
Thus the tensor force might have strong effects
on the extreme sub-barrier fusion, which is worth being
investigated in future.
\end{itemize}

\section*{Acknowledgments}
This work has been supported by
the National Natural Science Foundation of China (Grants No. 11975237,
No. 11575189, and No. 11790325) and
the Strategic Priority Research Program of Chinese Academy of Sciences
(Grant No. XDB34010000 and No. XDPB15).
Part of the calculations are performed on
the High-performance Computing Cluster of ITP-CAS and
the ScGrid of the Supercomputing Center,
Computer Network Information Center of Chinese Academy of Sciences.

\vspace*{2mm}

\end{document}